\documentstyle[preprint,aps,epsfig]{revtex}
\begin{document}
\draft
\preprint{LA PLATA $-$ TH 96/03}
\title{Approximate flavor symmetries at a grand unification scale}
\author{D.\ G\'omez Dumm\thanks{Fellow of CONICET (Argentina)}
\\Laboratorio de F\'{\i}sica Te\'orica, \\
Departamento de F\'{\i}sica, U.\ N.\ L.\ P.\ \\
c.\ c.\ 67, 1900 La Plata, Argentina.}

\maketitle

\begin{abstract}
We study the evolution of fermion mass matrices considering the hypothesis of
approximate flavor symmetries (AFS) in the Standard Model and a
two-Higgs-doublet model. We find that the hierarchical structure is not
significantly altered by the running, hence the assumption of AFS is entirely
compatible with a grand unification scenario.
\end{abstract}


The presence of strong hierarchies among fermion masses and mixing angles is
in general regarded as a signal of new physics beyond the Standard Model (SM).
At present, many different models have been developed, searching for a
mechanism which could give rise to the observed parameter hierarchies in a
``natural'' way.

In this work, we will concentrate on a model based on the existence of
approximate flavor symmetries (AFS), which has been recently proposed
by Antaramian, Hall and Ra\v{s}in \cite{rasin}. This model assumes the
presence of a set of global $U(1)$ symmetries (one for each quark
and lepton flavor) which are only slightly broken, conducing to different
suppression factors in the fermion mass matrices. The possible extension to
many Higgs doublets \cite{wei}, as well as the presence of neutrino
oscillations \cite{rasin2}, have also been studied within this scheme.

Our purpose is to test the possibility of inserting the idea of AFS into the
context of a grand unified theory (GUT), where the $SU(3)_C\otimes
SU(2)_L\otimes U(1)_Y$ gauge structure of the SM is nothing but the low-energy
manifestation of a theory containing a single gauge coupling constant. In such
a framework, it is natural to expect that the fermion mass matrix textures will
be determined at the scale of grand unification (typically of order $\sim
10^{15}$ GeV), rather than the electroweak one. That is what would happen if
the gauge and the $U(1)$ flavor symmetries were spontaneously broken at the
same time. The form of the mass matrices at the electroweak scale will then be
obtained by considering the evolution of the Yukawa couplings through the
renormalization group equations (RGE).

\hfill

\hfill

{\em Evolution of AFS with the energy scale}. In the Standard Model, the
Yukawa couplings can be written as
\begin{equation}
{\cal L}_Y=\sum_{i,j=1}^3 \left(
\lambda_{ij}^U\,\bar Q_iU_j\frac{\tilde{H}}{\sqrt{2}} +
\lambda_{ij}^D\,\bar Q_iD_j\frac{H}{\sqrt{2}} +
\lambda_{ij}^E\,\bar L_iE_j\frac{H}{\sqrt{2}} \right) +
\mbox{ h.c.}
\label{yuk}
\end{equation}
where $Q_i$ and $L_i$ are quark and lepton $SU(2)_L$ doublets and $U_j$, $D_j$
and $E_j$ stand for right fermion fields (we assume by now that only one
Higgs doublet $H$ is present). The hypothesis of AFS determines the order of
magnitude for the coupling parameters $\lambda$, leading to the relations
\begin{equation}
|\lambda^U_{ij}|\approx \epsilon_{Q_i}\epsilon_{U_j}\,,\;\;\;\;\;
|\lambda^D_{ij}|\approx \epsilon_{Q_i}\epsilon_{D_j}\,,\;\;\;\;\;
|\lambda^E_{ij}|\approx \epsilon_{L_i}\epsilon_{E_j}
\label{apro}
\end{equation}
where the $\epsilon_{F_i}$'s are dimensionless suppression factors originated
with the flavor symmetry breakdown. Their values can be approximately obtained
from the experimental data on quark masses and mixing angles.

In refs.\ \cite{rasin,wei,rasin2}, the relations (\ref{apro}) are assumed to
hold at the scale $Q^2\sim M_Z^2$. However, as stated above, the $\lambda$
matrices will evolve with energy. We are interested here in the form of the
Yukawa couplings at the large energy scale $M_X\sim 10^{15}$ GeV, hence we will
consider the one-loop RGE for the $\lambda$'s in the SM. These can be written
as \cite{lind}
\begin{equation}
\frac{d\lambda^F}{dt} = -\frac{1}{16\pi^2}\left[G_F(t)\openone-T_F(t)\openone-
\frac{3}{2}S_F(t)\right]\lambda^F
\label{rge}
\end{equation}
where $\openone$ and $S_F$ (as well as $\lambda^F$) are matrices in flavor
space, with $F=U,D,E$. We have used here the definitions
\begin{mathletters}
\label{defini}
\begin{eqnarray}
& & t=\frac{1}{2} \ln(Q^2/M_Z^2) \\
& & T_F=\mbox{Tr}\,(3\,\lambda^U\lambda^{U\dagger}+
3\,\lambda^D\lambda^{D\dagger}+\lambda^E\lambda^{E\dagger}) \\
& & S_U=-S_D=\lambda^U\lambda^{U\dagger}-\lambda^D\lambda^{D\dagger}\,,\;\;
S_E=\lambda^E\lambda^{E\dagger} \\
& & G_U=8\,g_3^2+\frac{9}{4}\,g_2^2+\frac{17}{12}\,g_1^2\,,\;\;
G_D=8\,g_3^2+\frac{9}{4}\,g_2^2+\frac{5}{12}\,g_1^2\,,\;\;
G_E=\frac{9}{4}\,g_2^2+\frac{15}{4}\,g_1^2
\end{eqnarray}
\end{mathletters}
being $g_1$, $g_2$ and $g_3$ the running coupling constants corresponding to
the $U(1)_Y$, $SU(2)_L$ and $SU(3)_C$ symmetry groups respectively. As it is
well known, their evolution is given by
\begin{equation}
\label{evolg}
g_i^2(t_2)=g_i^2(t_1)\left(1-\frac{K_i}{8\pi^2}\,
(t_2-t_1)\,g_1^2(t_1)\right)^{-1}
\end{equation}
with
\begin{equation}
K_1=\frac{20}{9}N+\frac{1}{6}\,,\;\;\;\;\;
K_2=\frac{4}{3}N-\frac{43}{6}\,,\;\;\;\;\;
K_3=\frac{4}{3}N-11
\end{equation}
($N=$ number of generations).

Let us write now the $\lambda$ matrices using their approximate form
(\ref{apro}), and ignoring for simplicity the presence of complex phases.
{}From (\ref{rge}), the $\epsilon$ parameters corresponding to
$\lambda^U$ will evolve with the energy scale according to
\begin{equation}
-16\pi^2\frac{d\;}{dt}\ln\left(\epsilon_{Q_i}\epsilon_{U_j}\right)=\alpha_U(t)
\label{evol}
\end{equation}
with
\begin{equation}
\alpha_U(t)\equiv G_U(t)-T_F(t)-\frac{3}{2}\sum_k\epsilon_{Q_k}^2
\sum_l\left(\epsilon_{U_l}^2-\epsilon_{D_l}^2\right)
\end{equation}
Similar expressions can be easily found for $F=D$ and $E$. In this way, at
the scale $t_X=1/2\ln(M_X^2/M_Z^2)$ we find
\begin{equation}
(\epsilon_{Q_i}\epsilon_{U_j})(t_X)=
(\epsilon_{Q_i}\epsilon_{U_j})(0)\,
\exp\left(-\frac{1}{16\pi^2}\int_0^{t_X}\!\alpha_U(t)\,dt\right)
\label{mult}
\end{equation}
This means that the running of $\lambda^U$ (and equivalently $\lambda^{D,E}$)
has not changed the ratios between the matrix elements: the sole effect has
been the introduction of a global multiplicative constant. On the other hand,
if AFS are assumed to hold at the large scale $t_X$ (i.e., just after the
spontaneous symmetry breakdown), they will be preserved when the coupling
constants run down to the electroweak energies. Thus the AFS
hypothesis, as presented in ref.\ \cite{rasin}, should in principle be
compatible with a grand unification scheme.

It is worth to notice that the equation (\ref{evol}) is just an approximate
result, since we have taken the relations (\ref{apro}) as if they were exact.
In fact, this would conduce to rank 1 quark and lepton mass matrices, that is,
to two zero mass eigenvalues for each fermion type. The corresponding (order
unity) corrections, however, will appear only in the last term of $\alpha_U$,
being suppressed by the small $\epsilon^2$ parameters.

\hfill

\hfill

{\em Numerical analysis.} Once the mass matrix elements are determined at the
scale $Q^2=M_Z^2$, the multiplicative constants $\beta_F\equiv
\exp(-1/(16\pi^2)\int^{t_X}_0\alpha_F\,dt)$ can be estimated numerically. For
definiteness, we will use the $\epsilon_{Q_i}/\epsilon_{Q_j}$ ratios given in
ref.\ \cite{wei} (we assume that they do not vary significantly when running
from 1 to 90 GeV), together with the quark mass values \cite{gasleu} at the
scale of the $Z$ mass \cite{lind}:
\begin{mathletters}
\label{masas}
\begin{equation}
\epsilon_{Q_1}/\epsilon_{Q_2}\approx 0.2\hspace{2cm}
\epsilon_{Q_2}/\epsilon_{Q_3}\approx 0.04
\end{equation}
\begin{equation}
m_d\simeq 5.6 \mbox{ MeV}\,,\;\;\;
m_s\simeq 0.11 \mbox{ GeV}\,,\;\;\;
m_b\simeq 3.4 \mbox{ GeV}
\end{equation}
\begin{equation}
m_u\simeq 3.2 \mbox{ MeV}\,,\;\;\;
m_c\simeq 0.85 \mbox{ GeV}
\end{equation}
\end{mathletters}
We will concentrate in the quark sector (the lepton matrix $\lambda^E$
approximately decouples in (\ref{evol})), ignoring as before the presence of
complex phases. The matrix elements $\lambda^{U,D}_{ij}$ can be approximately
calculated from the values in (\ref{masas}) taking into account the relations
\cite{wei}
\begin{equation}
\label{rela}
\epsilon_{Q_i}\epsilon_{F_i}\approx \frac{m_{F_i}}{\langle\phi\rangle}
\end{equation}
where $\langle\phi\rangle\simeq 175$ GeV is the vacuum expectation value (VEV)
of the neutral Higgs field.

Our numerical results are represented in figure 1, where the solid lines stand
for the values of $\beta_U$ and $\beta_D$ as functions of the top quark mass.
We also include a dashed line, which corresponds to the value of the matrix
element $\lambda^U_{33}$ at the scale
$M_X$. This parameter deserves special interest, since it is the largest of the
$\lambda^F_{ij}$ appearing in the Yukawa couplings. At the electroweak scale,
the relation $\lambda^U_{33}\simeq m_t/\langle\phi\rangle$ conduces to
$\lambda^U_{33}\approx 1$ for the expected range of $m_t$, hence the
identification of $\epsilon_{Q_3}$ and $\epsilon_{U_3}$ as {\em suppression}
factors is questionable. However, as it is shown in the figure, the situation
is considerably improved at the $M_X$ scale, where the value of
$\lambda^U_{33}$ lies between 1/3 and 1/2 for $m_t$ varying from 150 to 200
GeV. To this respect, notice that although the symbol ``$\approx$'' in
(\ref{rela}) indicates that the equalities are in general just approximate
(within a factor 2 or 3), the large value of $m_t$ compared to $m_{u,c}$
implies that the relation has to be almost exact for $F=U$, $i=3$.

\hfill

\hfill

{\em Two-Higgs-doublet case}. Let us finally carry out a similar analysis for a
two-Higgs-doublet model
(THDM) with natural flavor conservation (NFC) \cite{clas}. In this scheme, the
form of the Yukawa couplings will be entirely similar to that of eq.\
(\ref{yuk}), except for the replacements $H\rightarrow H_1$ and
$\tilde{H}\rightarrow\tilde{H}_2$. The NFC condition ensures that only one
Higgs couples to all the quarks of a given electric charge, preventing in this
way the presence of flavor-changing
processes at the tree level\footnote{It
has been argued that with the introduction of AFS, the NFC requirement could
be avoided. However, in that case further assumptions would be necessary to
suppress the appearing CP-violating phases \cite{wei}.}. Notice that one has
to deal with two non-zero Higgs VEVs, $v_1$ and $v_2$, therefore it is usual
to introduce a new parameter,
\begin{equation}
\tan \beta\equiv\frac{v_2}{v_1}
\end{equation}

As before, we will assume the AFS structure (\ref{apro}) and study the
evolution of the coupling ``constants'' with energy. The relevant one-loop RGE
for the THDM can be written as in (\ref{rge}) with the definitions \cite{lind2}
\begin{mathletters}
\begin{eqnarray}
& & S_U=\lambda^U\lambda^{U\dagger}
+\frac{1}{3}\lambda^D\lambda^{D\dagger}\,,\;\;\;
S_D=\frac{1}{3}\lambda^U\lambda^{U\dagger}+\lambda^D\lambda^{D\dagger} \\
& & T_U=\mbox{Tr}\,(3\,\lambda^U\lambda^{U\dagger})\,,\;\;\;
T_D=T_E=\mbox{Tr}\,(3\,\lambda^D\lambda^{D\dagger}+\lambda^E\lambda^{E\dagger})
\end{eqnarray}
\end{mathletters}
being $t$, $S_E$ and $G_F$ the same as in (\ref{defini}). The evolution of
the $g_i$'s has also the form (\ref{evolg}), where the constants $K_i$ are now
given by
\begin{equation}
K_1=\frac{20}{9}N+\frac{1}{3}\,,\;\;\;\;\;
K_2=\frac{4}{3}N-7\,,\;\;\;\;\;
K_3=\frac{4}{3}N-11
\end{equation}

In analogy with the Standard Model, it is immediate to find that the evolution
of the $\lambda$ matrices does not change the relations between the matrix
elements. We have for the THDM
\begin{equation}
\lambda^F_{ij}(t_X)=
\lambda^F_{ij}(0)\,\beta_F^{(2h)}=
\lambda^F_{ij}(0)\,
\exp\left(-\frac{1}{16\pi^2}\int_0^{t_X}\!\alpha^{(2h)}_F(t)\,dt\right)
\end{equation}
where
\begin{equation}
\alpha_{U,D}^{(2h)}(t)\equiv G_{U,D}(t)-T_{U,D}(t)-\frac{1}{2}\sum_k
\epsilon_{Q_k}^2\sum_l\epsilon_{{(U,D)}_l}^2
\end{equation}
Consequently, it is plausible to assume that the AFS are originated at a grand
unification scale, just like in the SM case.

In order to get numerical estimations for the $\beta_{U,D}^{(2h)}$ factors, we
will take once again the values in (\ref{masas}). However, due to the
two-doublet structure of the model, the relations (\ref{rela}) will now be
replaced by
\begin{equation}
\epsilon_{Q_i}\epsilon_{D_i}\approx \frac{m_{D_i}}{v_1}\hspace{2cm}
\epsilon_{Q_i}\epsilon_{U_i}\approx \frac{m_{U_i}}{v_2}
\end{equation}
while the VEVs have to satisfy the constraint
\begin{equation}
\left(v_1^2+v_2^2\right)^\frac{1}{2}\simeq 175 \mbox{ GeV}
\end{equation}
resulting from the electroweak symmetry breakdown. The value of $\tan \beta$
is a new unknown parameter, which could in principle be determined
experimentally.

In figure 2, we present our numerical results for $\beta_U^{(2h)}$ (a similar
behaviour is found for $\beta_D^{(2h)}$) and
$\lambda^U_{33}(M_X)$ as functions of $m_t$, for different values of
$\tan\beta$. It can be seen that if $\tan\beta\alt 1$, the value of
$\lambda_{33}^U$ is less than one only for $m_t\sim 150$ GeV, growing severely
as long as $m_t$ approaches 170 GeV. The regard of $\lambda^U_{33}$ as a
parameter reflecting an approximate flavor symmetry could be plausible only for
large values of $\tan\beta$, where the Yukawa couplings for the up-type quarks
are similar to those of the SM. Hence, taking into account the existing
phenomenological bounds on $\tan\beta$ \cite{libro}, the possibility of
assuming AFS at the $M_X$ scale within the THDM appears to be considerably
reduced in comparison with the Standard Model case.

\hfill

I would like to acknowledge Prof.\ L.\ Epele for a critical revision of the
manuscript. This work has been partially supported by Universidad Nacional
de La Plata (Argentina).

\pagebreak

\begin{figure}[htbp]
\begin{center}
\epsfig{file=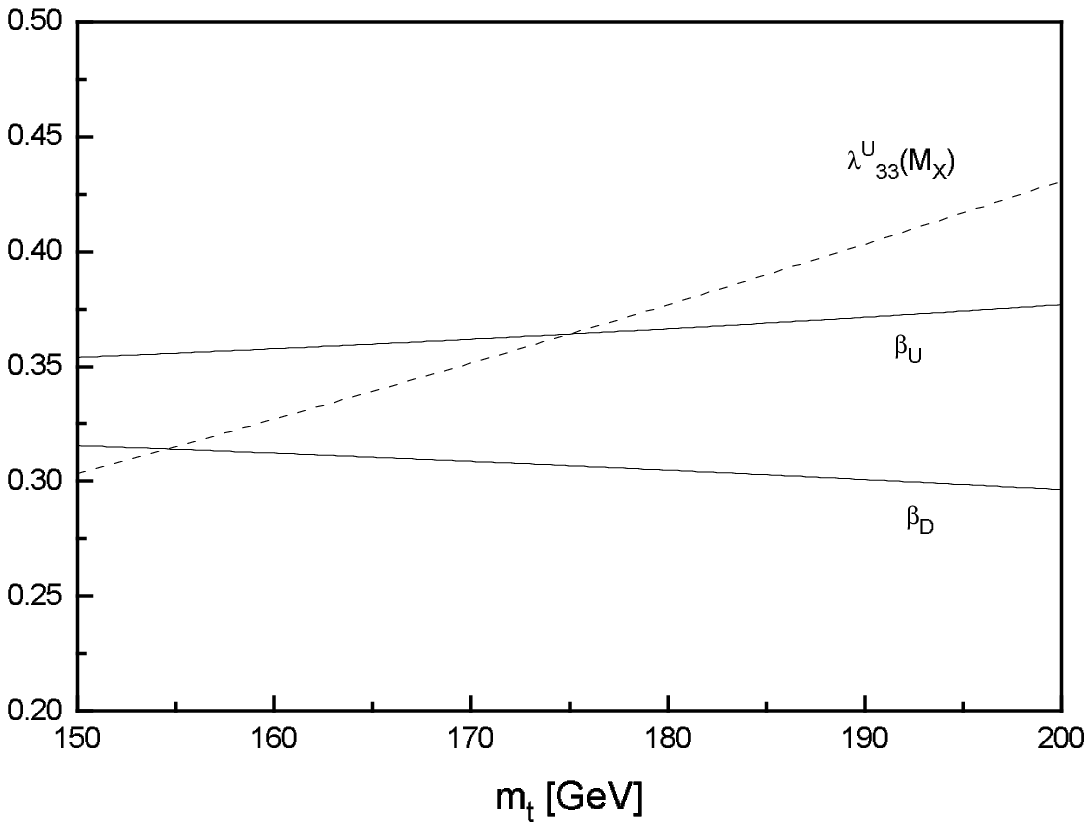}
\end{center}
\caption{Numerical results for the $\beta_{U,D}$ factors (solid lines) and
$\lambda^U_{33}(M_X)$ (dashed) as functions of the top quark mass.}
\end{figure}

\begin{figure}[htbp]
\begin{center}
\epsfig{file=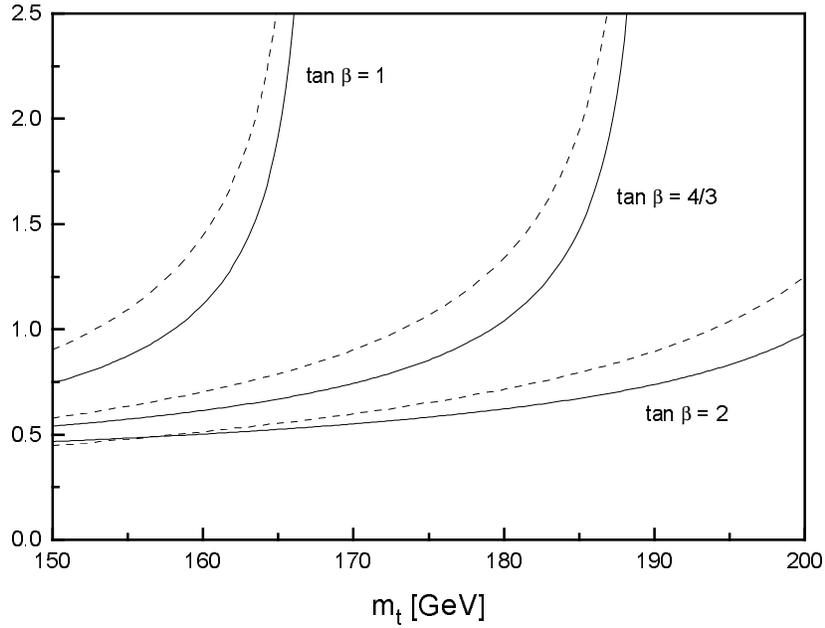}
\end{center}
\caption{Numerical results for the $\beta_U^{(2h)}$ factor (solid line)
and $\lambda^U_{33}(M_X)$ (dashed) in the THDM for different values of
$\tan\beta$ and the top quark mass.}
\end{figure}

\end{document}